\documentclass[journal=jpclcd,manuscript=letter]{achemso}

\usepackage{graphicx}
\usepackage{dcolumn}
\usepackage{bm}
\usepackage{amsmath}
\usepackage{amssymb}
\usepackage{siunitx}
\usepackage{makecell}
\usepackage{subfigure}
\usepackage{amsfonts}
\usepackage{multicol}
\usepackage{bm}
\usepackage{bbm}
\usepackage[version=4]{mhchem}
\usepackage{tabularx}
\usepackage{multirow}

\usepackage[normalem]{ulem}  

\newcommand{\ii}{\mathbbm{i}}
\newcommand{\tr}{\mathrm{tr}}

\newcommand{\hH}{\hat{H}}
\newcommand{\hV}{\hat{V}}
\newcommand{\hU}{\hat{U}}

\newcommand{\thalf}{\frac{T}{2}}

\author{Yuqi Wang}
\author{Wei-Hai Fang}
\author{Zhendong Li}
\email{zhendongli@bnu.edu.cn}
\affiliation{
Key Laboratory of Theoretical and Computational Photochemistry, Ministry of Education, College of Chemistry, Beijing Normal University, Beijing 100875, China
}

\title{Generalized many-body perturbation theory for the electron correlation energy: multi-reference random phase approximation via diagrammatic resummation}

\abbreviations{IR,NMR,UV}

\begin{document}

\begin{tocentry}

\includegraphics[width=\linewidth]{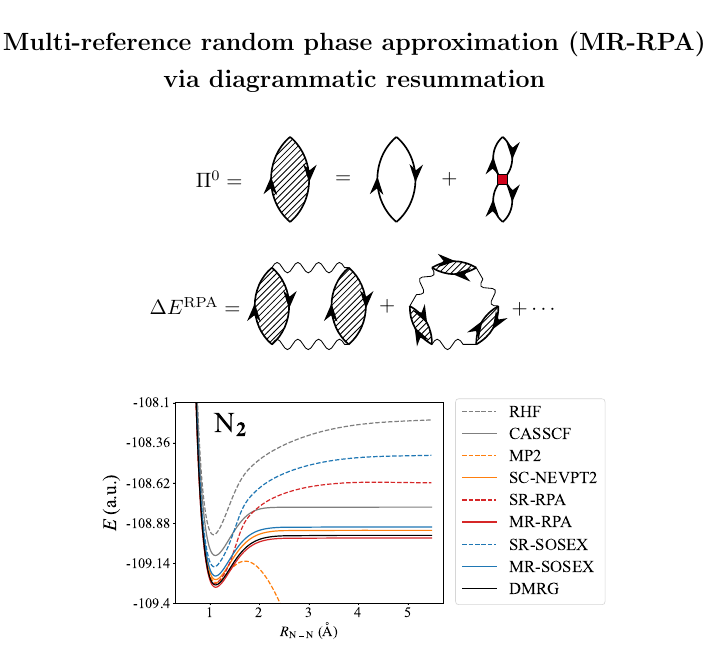}

\end{tocentry}

\begin{abstract}
Many-body perturbation theory (MBPT) based on Green's functions and Feynman diagrams 
provides a fundamental theoretical framework for various
\emph{ab initio} computational approaches in molecular and materials science, 
including the random phase approximation (RPA) and $GW$ approximation. Unfortunately, this perturbation expansion often  
fails in systems with strong multi-reference characters. 
Extending diagrammatic MBPT to the multi-reference case is highly nontrivial and remains largely unexplored, primarily due to the breakdown of Wick's theorem. 
In this work, we develop a diagrammatic multi-reference generalization of MBPT for computing correlation energies
of strongly correlated systems, by using the cumulant expansion of many-body Green's function in place of Wick's theorem. 
This theoretical framework bridges the gap between MBPT in condensed matter physics and multi-reference perturbation theories (MRPT) in quantum chemistry, which had been almost exclusively formulated within time-independent wavefunction frameworks prior to this work. Our formulation enables the explicit incorporation of strong correlation effects from the outset as in MRPT, while treating residual weak interactions through a generalized diagrammatic perturbation expansion as in MBPT. 
As a concrete demonstration, we formulate a multi-reference (MR) extension of the standard single-reference (SR) RPA
by systematically resumming generalized ring diagrams, which naturally leads to a unified set of equations applicable to both SR and MR cases. Benchmark calculations on prototypical molecular systems reveal that MR-RPA successfully resolves the well-known failure of SR-RPA in strongly correlated systems.
This theoretical advancement paves the way for advancing \emph{ab initio} computational methods through diagrammatic resummation techniques in future.
\end{abstract}

\textit{Introduction.} Electron correlation plays a pivotal role in determining the electronic properties of
molecules and materials materials. Understanding electron correlation within many-electron systems through \emph{ab initio} calculations
remains a fundamental challenge in modern quantum physics and chemistry\cite{martin2016interacting,coleman_introduction_2015,helgaker2013molecular},
simply because most of realistic systems cannot be solved exactly.
Many-body perturbation theory (MBPT) 
based on Green's functions and time-dependent Feynman diagrams
provides a rigorous and intuitive theoretical framework
to develop useful computational methods\cite{fetter_quantum_1971,negele_quantum_1998,li2019stochastic,bighin2023diagrammatic} for studying electron correlation beyond mean-field approximation.
Among them, random phase approximation (RPA)\cite{pines_collective_1952,bohm_collective_1953} represents one of the simplest methods, which includes certain type of Feynman diagrams to infinite order\cite{gell-mann_correlation_1957}. Originally developed for interacting
electron gas, RPA has nowadays been widely used to compute the electron correlation energy for molecules and materials\cite{hesselmann_random-phase_2011,ren_random-phase_2012,eshuis_electron_2012,chen_random-phase_2017}.
It has also aroused substantial theoretical interests for further improving the accuracy
by combining with density functional theory (DFT) via
the adiabatic connection fluctuation-dissipation (ACFDT) theorem\cite{langreth_exchange-correlation_1977,furche_developing_2008,toulouse_adiabatic-connection_2009,paier_hybrid_2010,heselmann_correct_2011}, including
exchange effects\cite{gruneis_making_2009,angyan_correlation_2011,bates_communication_2013,hummel_screened_2019} as well as single-excitation corrections\cite{ren_beyond_2011}, connecting
to coupled cluster (CC) theory\cite{scuseria_ground_2008,jansen_equivalence_2010,peng_equivalence_2013,scuseria_particle-particle_2013,berkelbach_communication_2018,quintero-monsebaiz_connections_2022,tolle_exact_2023}, and choosing different two-particle interaction channels\cite{van_aggelen_exchange-correlation_2013,van_aggelen_exchange-correlation_2014}. These developments have significantly deepened our understanding of electron correlation in complex systems.

Despite the remarkable success of methods based on standard MBPT in weakly correlated systems, they often perform miserably in the presence of strong electron correlation, such as stretching and breaking of chemical bonds, as highlighted in recent studies for RPA and $GW$ approximation\cite{tahir2019comparing,ammar2024can}.
This failure motivates us to develop a generalization of MBPT for 
systems with strong multi-reference (MR) characters.
The breakdown of standard MBPT can be attributed to the use of a single Slater determinant as the reference state and a quadratic mean-field Hamiltonian as the zeroth-order Hamiltonian $\hat{H}_0$.
Thus, our approach is to employ an interacting zeroth-order Hamiltonian as adopted
in many multi-reference perturbation theories (MRPT) in quantum chemistry,
such as the second-order $N$-electron valence perturbation theory (NEVPT2)\cite{angeli_introduction_2001}.
However, unlike traditional MRPT\cite{park2020multireference}, which are almost 
exclusively formulated within time-independent wavefunction frameworks,\cite{sokolov_time-dependent_2016}
generalizing diagrammatic MBPT to the multi-reference case in terms of Green's functions  
is nontrivial and remains largely unexplored, primarily due to the breakdown of Wick's theorem
for an interacting $\hat{H}_0$. 

In this work, we develop a diagrammatic multi-reference generalization of MBPT that bridges the gap between MBPT in condensed matter physics\cite{martin2016interacting,coleman_introduction_2015,helgaker2013molecular,negele_quantum_1998,stefanucci2013nonequilibrium} and MRPT in quantum chemistry\cite{park2020multireference}.
Our formulation enables the use of an interacting
$\hat{H}_0$ and a multi-determinant reference that treat strong correlation to infinite order from the outset
as in MRPT, while the residual weak interaction can be added by a generalized diagrammatic expansion similar to standard MBPT.
This framework is particularly useful for developing new theoretical approaches beyond second-order perturbation
theory. As a concrete demonstration, we formulate a multi-reference generalization of RPA and the second-order screened exchange (SOSEX)\cite{gruneis_making_2009} approximation. Compared with previous efforts on generalizing RPA\cite{yeager_multiconfigurational_1979,jorgensen_linear_1988,helmich-paris_casscf_2019,chatterjee_excitation_2012,pernal_intergeminal_2014, chatterjee_minimalistic_2016,pernal_electron_2018,pernal_exact_2018,pastorczak_correlation_2018,drwal_efficient_2022,matousek_toward_2023,guo_spinless_2024,szabados_ring_2017,margocsy_ring_2020}, our MR-RPA formulation has the distinct feature that it is derived from a clear diagrammatic resummation following the spirit of its single-reference counterpart by Gell-Mann and Brueckner\cite{gell-mann_correlation_1957}. Thus, it is systematically improvable by simply adding more diagrams as illustrated by MR-SOSEX. Benchmark calculations show that MR-RPA/SOSEX can successfully resolve the failure of SR-RPA/SOSEX, paving the way for developing more accurate computational methods via diagrammatic resummation.

\textit{Generalized many-body perturbation theory.} To solve the electronic Schr\"{o}dinger equation $\hat{H}|\Psi\rangle=E|\Psi\rangle$ using perturbation theory, we first introduce a partition of the second-quantized electronic Hamiltonian $\hH$, viz.,
\begin{eqnarray}
\hH &=& \hH_0 + \hV , \label{eq:partition}\\
\hH_0 &=& h_{pq}\hat{p}^\dagger \hat{q} + \frac{1}{2}h_{pr,qs}\hat{p}^\dagger\hat{q}^\dagger \hat{s}\hat{r} , \label{PT-H0}\\
\hV &=& v_{pq}\hat{p}^\dagger\hat{q} + \frac{1}{2}v_{pr,qs}\hat{p}^\dagger\hat{q}^\dagger\hat{s}\hat{r} ,\label{PT-V}
\end{eqnarray}
where the Einstein summation convention for repeated indices has been implied,
$h_{pq}$ ($h_{pr,qs}$) and $v_{pq}$ ($v_{pr,qs}$) are the zeroth-order and first-order one-electron (two-electron) interactions, respectively, and $\hat{p}^{(\dagger)}$ represents the Fermionic annihilation (creation) operator for the $p$-th spin-orbital. The sum of $h_{pr,qs}$ and $v_{pr,qs}$ obeys
$h_{pr,qs}+v_{pr,qs}=\langle pq|rs\rangle$, where $\langle pq|rs\rangle$ denotes the two-electron Coulomb integral\cite{martin2016interacting,helgaker2013molecular}. The partition \eqref{eq:partition} is very general in the sense that once the zeroth-order Hamiltonian $\hat{H}_0$ is specified, the perturbation operator is obtained as $\hat{V}=\hat{H}-\hat{H}_0$. In particular, standard MBPT corresponds to $h_{pr,qs}\equiv 0$, resulting in a quadratic $\hat{H}_0$ whose eigenstates
are Slater determinants. Here, we relax this restriction, allowing certain parts of $h_{pr,qs}$ corresponding to strong
electron interactions nonzero. This implies that the eigenfunction of $\hat{H}_0$ will be a multi-determinant
wavefunction in general, and the Wick's theorem\cite{wick1950evaluation} does not hold. The specific choice of $\hat{H}_0$ will be discussed later.

We assume that $|\Phi_0\rangle$ ($|\Psi_0\rangle$) and $E_0^{(0)}$ ($E_0$) are the non-degenerate ground-state wavefunction and energy
of $\hat{H}_0$ ($\hat{H}$), respectively. Then, the energy correction $\Delta E= E_0-E_0^{(0)}$ can be expressed by\cite{negele_quantum_1998} (see Supporting Information for details)
\begin{eqnarray}
\Delta E = \lim_{T\rightarrow \infty}\frac{\ii}{T}\ln \langle\hU(\frac{T}{2},-\frac{T}{2})\rangle_0,\label{eq:eshift}
\end{eqnarray}
where $\langle \hU(\frac{T}{2},-\frac{T}{2})\rangle_0$ is a shorthand notation for
$\langle\Phi_0|\hU(\frac{T}{2},-\frac{T}{2})|\Phi_0\rangle$, $\hat{U}(\frac{T}{2},-\frac{T}{2})=\mathcal{T}\exp\left(-\ii \int_{-\thalf}^{\thalf} \hat{V}(t)dt\right)$ with $\hat{V}(t)=
e^{\ii\hat{H}_0 t}\hat{V} e^{-\ii\hat{H}_0t}$, and $\mathcal{T}$ is the time-ordering operator. Throughout
this work, the time variable $t$ will be understood to be on a contour\cite{negele_quantum_1998} 
$\mathcal{C}=\{t\equiv(1-\ii\eta)\tilde{t}:\tilde{t}\in\mathbbm{R},\eta>0\}$ with $\eta$ being an infinitesimal parameter,
and the factor $-\ii\eta$ will only be explicitly included when necessary. Introducing an order counting parameter $\lambda$
for $\hat{V}$ in Eq. \eqref{eq:partition}, we can express $\langle\hU(\frac{T}{2},-\frac{T}{2})\rangle_0=
\sum_{n=0}\frac{\lambda^n}{n!}\mu_n$ where
\begin{eqnarray}
\mu_n = (-\ii)^n
\int_{-\thalf}^{\thalf}dt_1
\cdots
\int_{-\thalf}^{\thalf}dt_n 
\langle\mathcal{T}[\hat{V}(t_1)\cdots
\hat{V}(t_n)]\rangle_0.\label{eq:mun}
\end{eqnarray}
Interpreting $\langle\hU(\frac{T}{2},-\frac{T}{2})\rangle_0$
as the moment generating function for the moments $\mu_n$, 
then $\ln\langle\hU(\frac{T}{2},-\frac{T}{2})\rangle_0=\sum_{n=1}\frac{\lambda^n}{n!}\kappa_n$ is the corresponding cumulant generating function for the cumulants
$\kappa_n$. This allows to identify the $n$-th order energy correction $\Delta E_n$ from Eq. \eqref{eq:eshift} simply as
\begin{eqnarray}
\Delta E_n = \lim_{T\rightarrow \infty}\frac{\ii}{T n!}\kappa_n,\quad n\ge 1.\label{eq:en} 
\end{eqnarray}
The connection of the energy correction to cumulant reveals
the size-extensivity\cite{shavitt2009many} of $\Delta E_n$,
and can be viewed a generalization of the linked cluster theorem in standard MBPT\cite{goldstone_derivation_1957,li2019stochastic}.

\begin{figure}[t]
    \centering
    \includegraphics[width=0.45\textwidth]{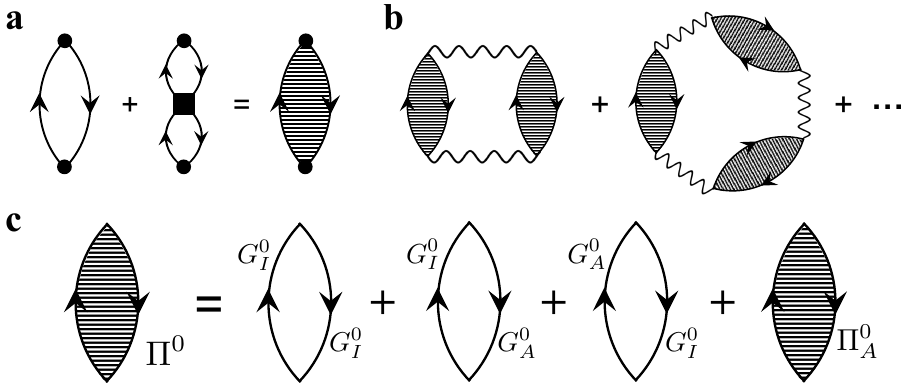}
     \caption{Generalized diagrammatic expansion. (a) Illustration of Second-order energy due to one-body perturbations (black dot). Each directed line connecting two interactions in the first diagram is a one-body Green's function $G_{pq}^0$. The black square along with two incoming and outgoing lines in the second diagram represent a connected two-body Green's function $G_{rs,pq}^{0,c}$ in Eq. \eqref{eq:cumulantGF}. 
     The sums of the two diagrams (without considering interactions) can be represented by the polarizability $\Pi_{pr,qs}^0$ (black bubble) in Eq. \eqref{eq:polarPi0}. (b) Definition of MR-RPA correlation energy in terms of a summation of generalized ring diagrams, where the polarizabilities defined in (a) are connected by the residual interaction (wiggle lines) $v_{pr,qs}$ in Eq. \eqref{PT-V}, rather than the bare Coulomb interaction $\langle pq|rs\rangle$ in SR-RPA.
     (c) Using the CASSCF reference and the Dyall Hamiltonian $\hat{H}_D$ in Eq. \eqref{eq:Hdyall-1main}, the polarizability can be separated into
     four parts, and only the part solely involving active orbitals contains
     a nonvanishing contribution from $G_{rs,pq}^{0,c}$.}\label{fig:method}
\end{figure}

To develop a generalized diagrammatic expansion for $\Delta E$,
we apply the cumulant decomposition\cite{negele_quantum_1998,metzner1991linked} of the time-ordered Green's functions in Eq. \eqref{eq:mun}.
To illustrate this more concretely, we consider the second-order energy correction due to one-body perturbations for simplicity. The integrand for $\kappa_2$ in this case involves 
\begin{equation*}
v_{pr}v_{qs}
\left(\langle\mathcal{T}
[\hat{p}^{\dagger}(t_1^+)\hat{r}(t_1)\hat{q}^{\dagger}(t_2^+)\hat{s}(t_2)]\rangle_0 - \langle\hat{p}^{\dagger}\hat{r}\rangle_0\langle \hat{q}^{\dagger}\hat{s}\rangle_0\right).
\end{equation*} We can identify
$\langle\mathcal{T}
[\hat{p}^{\dagger}(t_1^+)\hat{r}(t_1)\hat{q}^{\dagger}(t_2^+)\hat{s}(t_2)]\rangle_0$ as the zeroth-order two-body Green's function
$-G^0_{rs,pq}(t_1,t_2,t_1^+,t_2^+)$, and express it in terms of
connected Green's functions\cite{negele_quantum_1998,metzner1991linked}
\begin{eqnarray}
&&G^0_{rs,pq}(t_1,t_2,t_1^+,t_2^+) \nonumber\\
&=&
G^0_{rp}(t_1,t_1^+)G^0_{sq}(t_2,t_2^+) 
-G^0_{rq}(t_1,t_2^+)G^0_{sp}(t_2,t_1^+) \nonumber\\
&&+G^{0,c}_{rs,pq}(t_1,t_2,t_1^+,t_2^+),\label{eq:cumulantGF}
\end{eqnarray}
where $G^0_{pq}(t_1,t_2)=-\ii\langle\mathcal{T}[\hat{p}(t_1)\hat{q}^\dagger(t_2)]\rangle_0$ is the zeroth-order one-body Green's function 
and $G^{0,c}_{rs,pq}$ represents the
connected component of $G^0_{rs,pq}$,
which is nonzero for a generic $\hat{H}_0$
\eqref{PT-H0}. Then, the integrand becomes
$v_{pr}v_{qs}\left(G^0_{rq}(t_1,t_2^+)G^0_{sp}(t_2,t_1^+)-G^{0,c}_{rs,pq}(t_1,t_2,t_1^+,t_2^+)\right)$, 
which can be represented by two connected diagrams, see Fig. \ref{fig:method}(a).
Similar construction can be generalized
to arbitrary order for the partition \eqref{eq:partition}.
Consequently, analogous to the single-reference 
linked cluster theorem\cite{goldstone_derivation_1957}, 
we can rewrite Eq. \eqref{eq:en} as 
\begin{eqnarray}
\Delta E_n &=& \lim_{T\rightarrow \infty}\frac{\ii}{T}\frac{(-\ii)^n}{n!}
\int_{-\thalf}^{\thalf}dt_1
\cdots
\int_{-\thalf}^{\thalf}dt_n \nonumber\\
&&\quad\quad\langle\mathcal{T}[\hat{V}(t_1)\cdots 
\hat{V}(t_n)]\rangle_{0,\mathrm{linked}},\label{eq:enlinked}
\end{eqnarray}
where the subscript `linked' refers to \emph{generalized} linked diagrams, which in this case not only include standard Feynman diagrams (composed of interactions and $G^0_{pq}$), but also new diagrams involving zeroth-order connected many-body Green's functions such as $G^{0,c}_{rs,pq}$. In the special case that $\hat{H}_0$ is quadratic and $|\Phi_0\rangle$ is a Slater determinant, all connected many-body Green's functions vanish due to Wick's theorem\cite{wick1950evaluation} such that 
the present diagrammatic expansion reduces to standard MBPT using Feynman diagrams\cite{goldstone_derivation_1957}. 
To the best of our knowledge, the use of cumulant decomposition for many-body Green's functions in perturbation expansion first appeared in the linked-cluster expansion for the grand-canonical potential around the atomic limit of the Hubbard model\cite{metzner1991linked}, 
but it has not been applied in the context of \emph{ab initio} MBPT for calculating electron correlation energies before.

{\it Multi-reference random phase approximation.}
As a concrete application of the present theoretical framework, we use it to formulate a diagrammatic multi-reference generalization of RPA. Here, the term RPA refers to the direct RPA without exchange,
and the inclusion of exchange will be discussed later.
In the language of Feynman diagrams, the traditional single-reference RPA (SR-RPA) 
amounts to a resummation of ring diagrams
to infinite order\cite{bohm_collective_1953,gell-mann_correlation_1957}.
We can extend SR-RPA to the multi-reference case by simply defining MR-RPA as a summation of the generalized ring diagrams, see Fig. \ref{fig:method}(b),
where two interactions can be connected either by the conventional product
of two Green's functions $G^0_{rq}(t_1,t_2^+)G^0_{sp}(t_2,t_1^+)$
or by a connected two-body Green's function $G_{rs,pq}^{0,c}(t_1,t_2,t_1^+,t_2^+)$. 
While the former constitutes the conventional ring diagrams in SR-RPA,
the later also appears in the generalized diagrammatic expansion. 
As the local structure of the generalized ring diagrams is exactly the same as that in Fig. \ref{fig:method}(a) discussed before, we can rewrite the sum of these two contributions  
\begin{eqnarray}
&&G_{rq}^0(t_1,t_2^+)G_{sp}^0(t_2,t_1^+)
-
G_{rs,pq}^{0,c}(t_1,t_2,t_1^+,t_2^+) \nonumber\\
&=&
\langle \mathcal{T}\left[
\left(\hat{p}^\dagger(t_1^+) \hat{r}(t_1)
-\langle \hat{p}^\dagger \hat{r}\rangle_0\right)
\left(\hat{q}^\dagger(t_2^+) \hat{s}(t_2)-
\langle \hat{q}^\dagger \hat{s}\rangle_0\right)\right]\rangle_0\nonumber\\
&\equiv& \ii\Pi^0_{pr,qs}(t_1,t_2), \label{eq:polarPi0}
\end{eqnarray}
in terms of the irreducible polarizability\cite{stefanucci2013nonequilibrium} (polarization propagator) $\Pi^0_{pr,qs}(t_1,t_2)$.
Then, the MR-RPA correlation energy defined in Fig. \ref{fig:method}(b) can be expressed as
\begin{eqnarray}
\Delta E^{\mathrm{RPA}}
\equiv
\sum_{n \ge 2}
\Delta E_n^{\mathrm{ring}},
\end{eqnarray}
where the $n$-th order ring contribution in Eq. \eqref{eq:enlinked} is (see Supporting Information for details)
\begin{eqnarray}
\Delta E_n^{\mathrm{ring}}=
-\frac{1}{2n}\frac{1}{2\pi} \int_{-\infty}^{+\infty}d\omega
\,\tr([\mathbf{v}\mathbf{\Pi}^0(\ii\omega)]^n),
\end{eqnarray}
with $v_{pr,qs}$ given in Eq. \eqref{PT-V}. The diagrammatic expansion can be resummed into
\begin{eqnarray}
\Delta E^{\mathrm{RPA}}
=\int_{-\infty}^{+\infty}
\frac{d\omega}{2\pi}
\frac{1}{2}\tr\left[\ln[\mathbf{I}-\mathbf{v}\mathbf{\Pi}^0(\ii\omega)]+\mathbf{v}\mathbf{\Pi}^0(\ii\omega)\right]\label{Ecorr-log},
\end{eqnarray}
which is exactly the same as that for SR-RPA\cite{furche_developing_2008,ren_random-phase_2012,heselmann_random-phase-approximation_2012}. Thus, our MR-RPA can be viewed as a natural and seamless extension of SR-RPA 
to the multi-reference case with more general two-electron perturbation and irreducible polarizability.

{\it Dyson equation and plasmon formula.} While Eq. \eqref{Ecorr-log} can already be used to compute MR-RPA correlation energies via
numerical quadratures\cite{eshuis2010fast,ren2012resolution},
we derive an equivalent form in terms of an MR-RPA generalized eigenvalue problem
by integrating the frequency analytically\cite{furche_developing_2008}, 
which offers deeper theoretical insights into the MR-RPA method.
To this end, we introduce a reducible polarizability in
the form of a Dyson equation
\begin{eqnarray}
    \mathbf{\Pi}^{\lambda}(z)
\equiv
\mathbf{\Pi}^0(z)+\lambda\mathbf{\Pi}^0(z)\mathbf{v}\mathbf{\Pi}^\lambda(z),\label{eq:dysonPi}
\end{eqnarray}
where $z$ is a complex frequency and $\lambda$ is the order parameter
associated with $\hat{V}$. Then, Eq. \eqref{Ecorr-log} can be rewritten as a coupling constant integration
\begin{align}
    \Delta E^{\mathrm{RPA}}=
-\frac{1}{2}\int_0^1 d\lambda \int_{-\infty}^{\infty}\frac{d\omega}{2\pi}
\tr(\mathbf{v}[\mathbf{\Pi}^\lambda(\ii\omega)
-\mathbf{\Pi}^0(\ii\omega)
]). \label{Ecorr-double-int}
\end{align}
Using the spectral representation of
$\Pi^0_{pr,qs}(z)
=    
\sum_{N>0}\left[
\frac{\langle 0|\hat{p}^\dagger\hat{r}|N\rangle \langle N|\hat{q}^\dagger\hat{s}|0\rangle}
{z - \omega_{N}}
-
\frac{\langle 0|\hat{q}^\dagger\hat{s}|N\rangle \langle N|\hat{p}^\dagger\hat{r}|0\rangle }
{z + \omega_{N}}\right]$, where the ground and excited states of $\hat{H}_0$ are abbreviated by $|0\rangle$ and $|N\rangle$, respectively,
and $\omega_N=E_{N}^{(0)}-E_0^{(0)}$ is
the zeroth-order excitation energy,
both integrations in Eq. \eqref{Ecorr-double-int} can be performed analytically by extending the previous technique developed in 
SR-RPA\cite{furche_developing_2008} (see Supporting Information for details).
The final result, usually referred to as the plasmon formula in SR-RPA\cite{furche_developing_2008}, reads
\begin{eqnarray}
    \Delta E^{\mathrm{RPA}} = \frac{1}{2}\sum_{I>0}(\Omega_I^{\mathrm{RPA}}-\Omega_I^{\mathrm{TDA}})\label{eq:plasmonFormula},
\end{eqnarray}
where $I>0$ means that only positive $\Omega$ is concerned. Here, $\Omega_I^{\mathrm{RPA}}$ is given
by a generalized eigenvalue problem
\begin{eqnarray}
    \begin{bmatrix}
\mathbf{A} & \mathbf{B}\\
\mathbf{B}^* & \mathbf{A}^*
\end{bmatrix}
\begin{bmatrix}
\mathbf{x}_I\\
\mathbf{y}_I
\end{bmatrix}
=\begin{bmatrix}
    \mathbf{I} & \mathbf{0}\\
    \mathbf{0} & -\mathbf{I}
\end{bmatrix}
\begin{bmatrix}
\mathbf{x}_I\\
\mathbf{y}_I
\end{bmatrix}\Omega_I, \label{eq:casida}
\end{eqnarray}
where matrices $\mathbf{A}$ ($=\mathbf{A}^\dagger$) and $\mathbf{B}$ ($=\mathbf{B}^T$) are defined by
\begin{align}
A_{NM}&=\omega_N\delta_{NM} + \langle N|\hat{p}^\dagger \hat{r}|0\rangle v_{pr,qs} \langle 0|\hat{q}^\dagger \hat{s}|M\rangle,\label{casida-A}\\
B_{NM}&=\langle N|\hat{p}^\dagger \hat{r}|0\rangle v_{pr,qs} \langle M|\hat{q}^\dagger \hat{s}|0\rangle, \label{casida-B}
\end{align}
respectively, with $N,M>0$ labeling excited states of $\hat{H}_0$ and the Einstein
summation convention assumed for repeated orbital indices. 
$\Omega_I^{\mathrm{TDA}}$ corresponds to the eigenvalue
of Eq. \eqref{eq:casida} with $\mathbf{B}$ set to zero, often called 
the Tamm-Dancoff approximation (TDA) in SR-RPA\cite{Tamm:1945qv,PhysRev.78.382}. 
For quadratic $\hat{H}_0$, Eq. \eqref{eq:casida} reduces to the SR-RPA
equation, since $\langle N|\hat{p}^\dagger\hat{r}|0\rangle$
is nonzero only if $|N\rangle = \hat{p}^\dagger\hat{r}|0\rangle$
with $r$ and $p$ referring to occupied and virtual orbitals, respectively. Therefore, Eqs. \eqref{eq:plasmonFormula}-\eqref{casida-B} 
seamlessly bridge RPA in the SR and MR cases.

{\it Multi-reference second-order screened exchange correction.} 
So far, MR-RPA is developed without considering exchange, and hence it
will suffer from self-interaction error (SIE) as in 
SR-RPA\cite{henderson2010connection,mori-sanchez_failure_2012,tahir2019comparing}.
A simple way to include exchange is to generalize the
second-order screened exchange (SOSEX) correction\cite{gruneis_making_2009} 
to the multi-reference case. Specifically, the connection between Eq. \eqref{eq:casida} for SR-RPA and the direct ring coupled cluster doubles (drCCD)
equation 
\begin{eqnarray}
\mathbf{B}^* + \mathbf{A}^*\mathbf{T} + \mathbf{T} \mathbf{A}
+\mathbf{TBT} = \mathbf{0},\label{riccati}
\end{eqnarray}
has been established\cite{scuseria_ground_2008}, where $\mathbf{T}=\mathbf{YX}^{-1}$ with
$\mathbf{X}$ and $\mathbf{Y}$ being the eigenvectors of Eq. \eqref{eq:casida}
with positive eigenvalues. The SR-SOSEX\cite{gruneis_making_2009} amounts to evaluating Eq. \eqref{eq:plasmonFormula} by using its equivalent form
\begin{eqnarray}
\Delta E^{\mathrm{RPA}}=\frac{1}{2}\tr(\mathbf{BT}),\label{energy-drCCD}
\end{eqnarray}
but with $v_{pr,qs}$ in $\mathbf{B}$ replaced by its antisymmetrized counterpart $v_{pr,qs}-v_{ps,qr}$. Since our MR-RPA equation has exactly the same mathematical form, the similar derivation leads naturally to MR-SOSEX (see Supporting Information for details).

{\it Comparison with other generalizations of RPA.} 
We have presented three equivalent formulae for MR-RPA correlation energies, i.e.
Eqs. \eqref{Ecorr-log}, \eqref{eq:plasmonFormula}, and \eqref{energy-drCCD}, from three different perspectives. A few remarks on the relationship between the present work and previous MR generalizations\cite{yeager_multiconfigurational_1979,jorgensen_linear_1988,helmich-paris_casscf_2019,chatterjee_excitation_2012,pernal_intergeminal_2014, chatterjee_minimalistic_2016,pernal_electron_2018,pernal_exact_2018,pastorczak_correlation_2018,drwal_efficient_2022,matousek_toward_2023,guo_spinless_2024,szabados_ring_2017,margocsy_ring_2020}
of RPA are in order:

Firstly, while in the single reference case, different theoretical formulations of RPA,
including ring (or bubble) diagram resummation\cite{gell-mann_correlation_1957},
equation of motion (EOM)\cite{rowe1968equations},
time-dependent Hartree-Fock\cite{ODDERSHEDE1978275}, adiabatic connection\cite{langreth_exchange-correlation_1977,furche_density_2001}, and
ring approximation to coupled cluster doubles\cite{scuseria_ground_2008}, 
all give rise to equivalent results, they generally lead to different approximations 
in the multi-reference case, manifesting the complexity of the strong correlation problem.
For instance, the MR-RPA generalized eigenvalue problem with matrix elements given
in Eqs. \eqref{casida-A} and \eqref{casida-B} does not seem
to correspond to any equation derived from the EOM\cite{yeager_multiconfigurational_1979,jorgensen_linear_1988,helmich-paris_casscf_2019,chatterjee_excitation_2012,pernal_intergeminal_2014, chatterjee_minimalistic_2016,pernal_electron_2018,pernal_exact_2018,pastorczak_correlation_2018,drwal_efficient_2022,matousek_toward_2023,guo_spinless_2024}, although they share the similar form \eqref{eq:casida}.

Secondly, our generalization is more natural generalization of SR-RPA, in the sense that 
Eqs. \eqref{Ecorr-log}, \eqref{eq:plasmonFormula}, and \eqref{energy-drCCD}
share exactly the same mathematical forms as in the SR case.
In contrast, other MR generalizations \cite{yeager_multiconfigurational_1979,jorgensen_linear_1988,helmich-paris_casscf_2019,chatterjee_excitation_2012,pernal_intergeminal_2014, chatterjee_minimalistic_2016,pernal_electron_2018,pernal_exact_2018,pastorczak_correlation_2018,drwal_efficient_2022,matousek_toward_2023,guo_spinless_2024,szabados_ring_2017,margocsy_ring_2020}
of RPA do not have such simplicity and consistency. For instance, in the work by
Pernal et al.\cite{pernal_electron_2018,pernal_exact_2018,pastorczak_correlation_2018,drwal_efficient_2022,matousek_toward_2023,guo_spinless_2024}, the extended RPA equation derived
from the EOM is used to compute the transition density matrix,
while the correlation energy is derived from adiabatic connection
with additional approximation. Thus, there are no direct analogs of Eq. \eqref{Ecorr-log}, \eqref{eq:plasmonFormula}, or \eqref{energy-drCCD} in their case.

Thirdly, we observe that there is no direct correspondence between the MR-RPA equation \eqref{riccati} and existing multi-reference coupled-cluster (MRCC) theories\cite{lyakh2012multireference}. Thus, it differs from the multi-reference ring coupled-cluster approximations developed by Szabados and coworkers\cite{szabados_ring_2017,margocsy_ring_2020}. 
Potentially, it serves as a promising starting point for deriving novel MRCC theories.

{\it Choice of zeroth-order Hamiltonian.}
The above MR-RPA and MR-SOSEX are general in the sense that
they can be applied to any partition of form \eqref{eq:partition}.
However, to be both computationally feasible and accurate, the choice of zeroth-order Hamiltonian is crucial. In this work, we employ the complete active space self-consistent field\cite{helgaker2013molecular} (CASSCF) wavefunction as the zeroth-order state, which is the usual starting point of multi-reference perturbation theory in quantum chemistry\cite{park2020multireference}, and the Dyall Hamiltonian\cite{dyall_choice_1995} as $\hH_0$ among other possible partitions\cite{rosta_two-body_2002,fink_multi-reference_2009}. The Dyall Hamiltonian $\hH_D$ contains two parts
(see Supporting Information for detailed definitions of
$\epsilon_i$, $\epsilon_a$ and $f_{xy}$)
\begin{eqnarray}
    \hH_D &=& \hH_I+\hH_A ,\label{eq:Hdyall-1main}\\
    \hH_I &=& \epsilon_i \hat{i}^{\dagger}\hat{i} + \epsilon_a \hat{a}^{\dagger}\hat{a} ,\label{eq:Hdyall-2main}\\
    \hH_A &=& f_{xy}\hat{x}^{\dagger}\hat{y} + \frac{1}{2}
    \langle xy|zw\rangle \hat{x}^{\dagger}\hat{y}^{\dagger}\hat{w}\hat{z},\label{eq:Hdyall-3main}
\end{eqnarray}
where the so-called inactive part $\hH_I$ is quadratic for the doubly occupied and virtual orbitals, while the active part $\hH_A$ retains the full two-electron Coulomb interactions among the $M_A$ active orbitals, within which the electron correlation is supposed to be stronger. Due to its additive separability, all of its eigenfunctions (e.g., the CASSCF ground state $|0\rangle=|\Theta_0\rangle|\Phi_0^{N_A}\rangle$)
are products of an inactive part ($|\Theta_0\rangle$), which is simply a Slater determinant, and an active part ($|\Phi_0^{N_A}\rangle$), which is a multi-determinant wavefunction that describes the strong correlation among the $N_A$ active electrons within the active space denoted by CAS($N_A,M_A$). Consequently, the zeroth-order excited state $|N\rangle$ ($N>0$) that can couple with $|0\rangle$ in Eqs. \eqref{casida-A} and \eqref{casida-B} must belong to the
following four classes of excitations (see Fig. S2 in Supporting Information)
\begin{align}
\{|\Theta_i^a\rangle|\Phi_0^{N_A}\rangle, |\Theta_i\rangle|\Phi_\mu^{N_A+1}\rangle, |\Theta^a\rangle|\Phi_\mu^{N_A-1}\rangle, |\Theta_0\rangle|\Phi_{\mu>0}^{N_A}\rangle\},  \label{eq:se}
\end{align}
where $|\Theta_i^a\rangle=\hat{a}^\dagger\hat{i}|\Theta_0\rangle$,
$|\Theta_i\rangle=\hat{i}|\Theta_0\rangle$,
$|\Theta^a\rangle=\hat{a}^\dagger|\Theta_0\rangle$,
and $|\Phi_{\mu>0}^{N_A}\rangle$ represents the $\mu$-th excited state
of $\hat{H}_A$ with $N_A$ electrons.
The first class of excitations, analogous to those in SR-RPA,
is composed of single excitations from doubly occupied orbitals
to virtual orbitals, while the other three classes 
potentially involve multiple excitations within the active space.
The fourth class, in particular, consists of excitations within the active space.
All of the four classes contribute to the screening of two-electron interactions
at the MR-RPA level, see Fig. \ref{fig:method}(c), which goes beyond the conventional SR-RPA screening by particle-hole excitations only. In this regard, including the connected two-body Green's function $\langle\mathcal{T}[\hat{p}^\dagger(t_1^+) \hat{r}(t_2) \hat{q}^\dagger(t_2^+) \hat{s}(t_2)]\rangle_{0,c}$ into the irreducible polarizability
$\Pi^0_{pr,qs}(t_1,t_2)$ 
can be viewed as a special type of vertex correction solely
due to the two-electron interactions within
the active space (see Eq. \eqref{eq:Hdyall-3main}),
which is supposed to be the most essential for strong correlation.
Detailed matrix elements of the resulting 4-by-4 block matrices
($\mathbf{A}$ and $\mathbf{B}$) are presented
in Supporting Information.

\begin{figure}[t]
\includegraphics[width=0.46\textwidth]{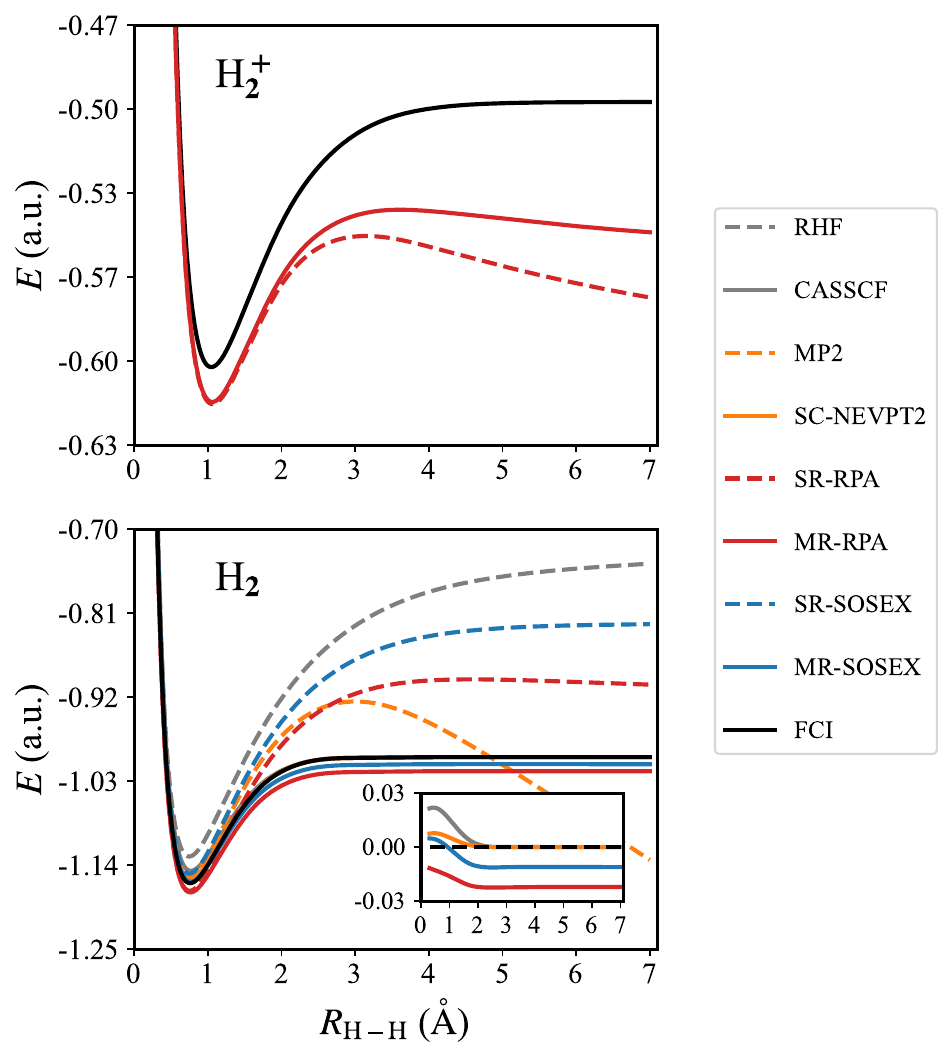}
\caption{Bond dissociation of \ce{H2+} and \ce{H2} 
calculated by different methods using the cc-pVDZ basis
set. The active spaces used for \ce{H2+} and \ce{H2}
are CAS(1,1) and CAS(2,2), respectively. 
For \ce{H2+}, other methods including RHF, MP2, CASSCF, SR-SOSEX, and MR-SOSEX are all exact, and hence not displayed for clarity. The inset in the second figure shows the errors of
multi-reference results with respect to the FCI result.
}\label{fig: H2 and H2+}
\end{figure}

\begin{figure*}[t]
    \centering
    \includegraphics[width=0.98\linewidth]{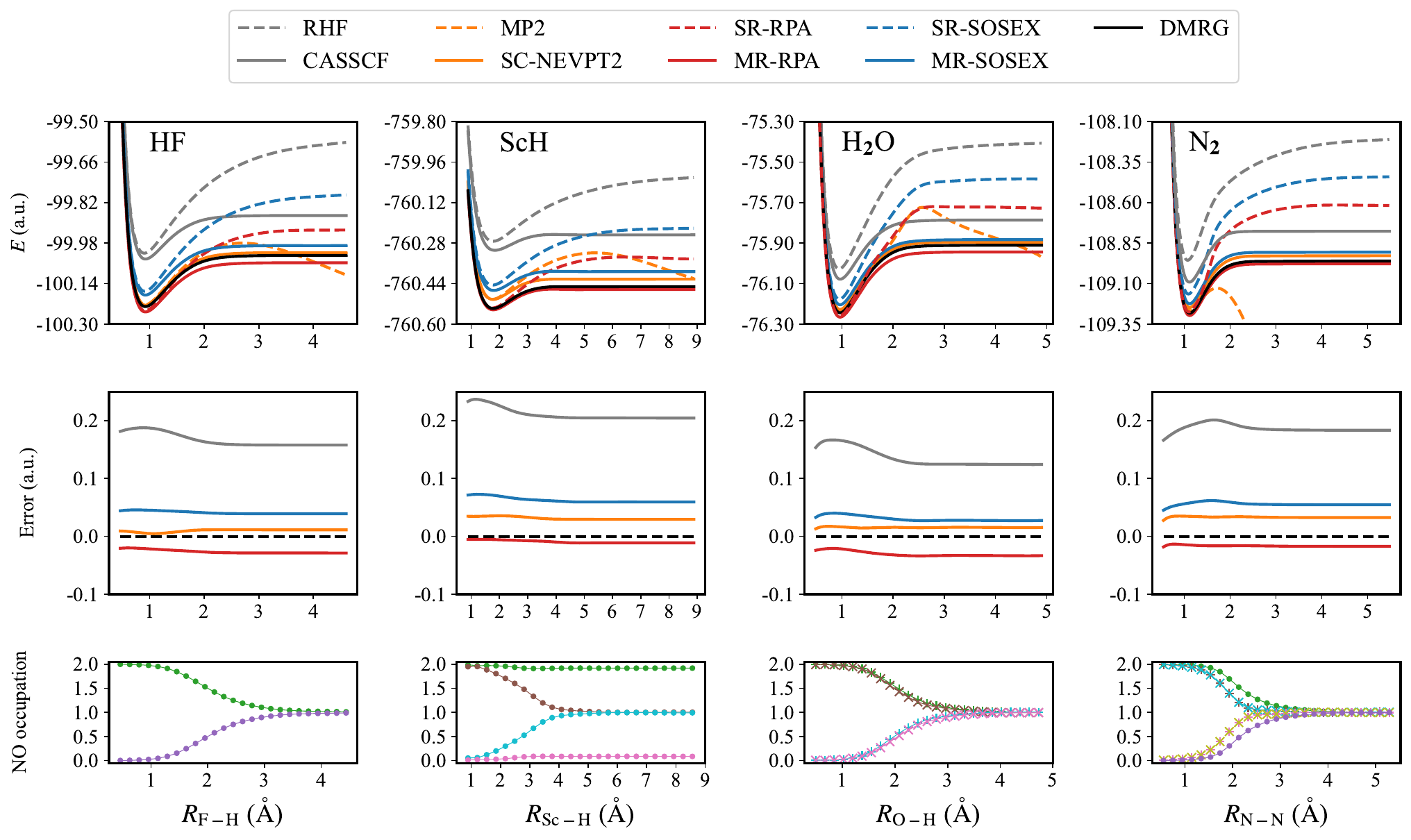}
    \caption{Potential energy curves for four molecules (\ce{HF}, \ce{ScH}, \ce{H2O} and \ce{N2}) computed with different methods using the cc-pVDZ basis set. Errors of multi-reference results with respect to the nearly exact DMRG results are shown in the middle. 
    The active spaces are CAS(2,2), CAS(4,4), CAS(4,4), and CAS(6,6) for \ce{HF}, \ce{ScH}, \ce{H2O} and \ce{N2}, respectively, with the CASSCF
    natural orbital (NO) occupation numbers shown below . Significant deviations from 0 or 2 indicate the presence of strong electron correlation. }
    \label{fig:HF-ScH-H2o-N2}
\end{figure*}

{\it Results and discussion.}

We have implemented MR-RPA and MR-SOSEX using Eqs. \eqref{eq:plasmonFormula} and \eqref{energy-drCCD}, respectively,
based on the PySCF package\cite{sun_recent_2020} (see Supporting Information for details).
Thus, the current implementation works for small molecules, for which the diagonalization of the MR-RPA eigenvalue problem
\eqref{eq:casida} is feasible.
More efficient implementation can be developed by computing the correlation energy based on Eq. \eqref{Ecorr-log}
using numerical quadratures and resolution of identity (RI) approximation\cite{eshuis2010fast,ren2012resolution},
which will be a subject of future study. In this work, we primarily focus on studying the performances of MR-RPA/MR-SOSEX for typical molecules.

Figure \ref{fig: H2 and H2+} shows the potential energy curves (PECs)
computed by various SR and MR methods using the cc-pVDZ basis set\cite{dunning_gaussian_1989} for \ce{H2+} and \ce{H2}, which have become a testbed for RPA\cite{mori-sanchez_failure_2012,bates_communication_2013,van_aggelen_exchange-correlation_2014}. For \ce{H2+}, both SR-RPA and MR-RPA suffer from the SIE mentioned before, 
resulting a lower energy than full configuration interaction (FCI)  
due to the lack of exchange in direct RPA. The SIE in MR-RPA is smaller, because the two-electron interactions in $\hat{H}_A$ \eqref{eq:Hdyall-3main} are treated exactly with exchange included to all orders.
In comparison, both SR-SOSEX and MR-SOSEX solve the SIE problem in direct RPA for \ce{H2+}.
In the case of \ce{H2}, the PEC obtained from second order M{\o}ller-Plesset perturbation theory (MP2) starting from a restricted Hartree-Fock (RHF) reference diverges as the bond distance increases.
Although SR-RPA and SR-SOSEX do not show such divergence, they still produce significant errors in this regime.
Compared with the exact FCI reference, the improvements from MR-RPA and MR-SOSEX are dramatic.
The non-parallel errors (NPEs) of MR-RPA and MR-SOSEX are 8.79 and 13.30 milli-Hartree (mH), respectively, which are much smaller than the NPE of CASSCF (18.27 mH) and are slightly larger than that (6.44 mH) of 
the strongly contracted version of NEVPT2 (SC-NEVPT2)\cite{angeli_introduction_2001}.

Figure \ref{fig:HF-ScH-H2o-N2} displays the PECs of four prototypical molecules (\ce{HF}, \ce{ScH}, \ce{H2O}, and \ce{N2}). 
As demonstrated by the occupation numbers of the CASSCF natural orbitals (NOs) within the active space, the electron correlation is progressively stronger from \ce{HF} to \ce{N2}. 
Again, we found that both SR-RPA and SR-SOSEX fail miserably at stretched geometries, indicating the break down of standard MBPT based on a RHF reference. 
At short bond distance, where electron correlation is weak and
the RHF and CASSCF energies are close to each other, the performances of our MR generalizations are consistent with those of 
their SR counterparts. However, at larger bond distance, MR-RPA/SOSEX improve their SR counterparts significantly
by treating strong correlation among active orbitals at the zeroth order. 
This is particularly important for \ce{N2}, where the SR-RPA/SOSEX curves deviate considerably from the corresponding MR-RPA/SOSEX results already around the equilibrium geometry. Note that since CASSCF completely misses the weak dynamic correlation, the CASSCF PECs are much higher than the nearly exact results obtained by using the \emph{ab initio} density matrix renormalization group (DMRG) algorithm\cite{PhysRevLett.69.2863,annurev-physchem-032210-103338,xiang2024distributed}.
In contrast, MR-RPA and MR-SOSEX treat both strong and weak electron correlation in a more balanced way,
and hence lead to a good agreement with the DMRG results for realistic molecules. 
As shown in Fig. \ref{fig:HF-ScH-H2o-N2}
and Table  \ref{tab: Edis and req} for
bond dissociation energies $\Delta E$,
equilibrium bond distances $R_{\mathrm{eq}}$,
and NPEs, the performances of MR-RPA, MR-SOSEX, and SC-NEVPT2 
are comparable. A more detailed theoretical comparison of 
MR-RPA and NEVPT2 will be presented elsewhere.

\begin{table}
    \setlength{\tabcolsep}{1.5mm}
    \caption{
    Comparison of different multi-reference methods for
    dissociation energies $\Delta E$, equilibrium bond distances $R_{\mathrm{eq}}$, and non-parallel errors (NPEs) with respect to the nearly exact DMRG results 
    using the cc-pVDZ basis set.
    }
    
    \centering
    \scalebox{0.8}{\begin{tabular}{ccccc}
    \hline\hline
    Molecule & Method &$ \Delta E$ (mH) & $R_{\textrm{eq}}$ (\AA) &NPE (mH)\\
    \hline
       \multirow{5}{*}{\ce{HF}} & CASSCF    & 171.8 & 0.920 & 29.97  \\
                                & MR-RPA   & 194.3 & 0.922  & 8.78  \\ 
                                & MR-SOSEX  & 195.4 & 0.920 & 6.76  \\ 
                                & SC-NEVPT2 & 207.8 & 0.923 & 6.55  \\ 
                                & DMRG      & 201.8 & 0.919 & /    \\ \hline
       \multirow{5}{*}{\ce{ScH}} & CASSCF   & 61.1 & 1.836 & 32.24 \\
                                & MR-RPA    & 80.4 & 1.781 &  6.20 \\ 
                                & MR-SOSEX  & 74.5 & 1.794 & 13.31\\ 
                                & SC-NEVPT2 & 80.2 & 1.773 &  6.17 \\ 
                                & DMRG      & 86.2 & 1.776 & /    \\ \hline
       \multirow{5}{*}{\ce{H2O}} & CASSCF   & 292.0 & 0.966 & 42.38 \\
                                & MR-RPA   & 321.7 & 0.965 & 12.85 \\ 
                                & MR-SOSEX  & 321.3 & 0.965 & 12.72 \\ 
                                & SC-NEVPT2 & 332.5 & 0.965 & 4.20  \\ 
                                & DMRG      & 333.7 & 0.963 & /    \\ \hline
       \multirow{5}{*}{\ce{N2}} & CASSCF    & 313.9 & 1.114 & 35.05 \\
                                & MR-RPA   & 319.6 & 1.120 & 4.93  \\ 
                                & MR-SOSEX  & 318.6 & 1.117 & 16.58 \\ 
                                & SC-NEVPT2 & 319.7 & 1.120 & 7.45  \\ 
                                & DMRG     & 321.8 & 1.119 & /    \\ 
    \hline\hline
    \end{tabular}}


    \label{tab: Edis and req}
\end{table}

\begin{figure}[t]
    \centering
    {
    \includegraphics[width=.48\textwidth]{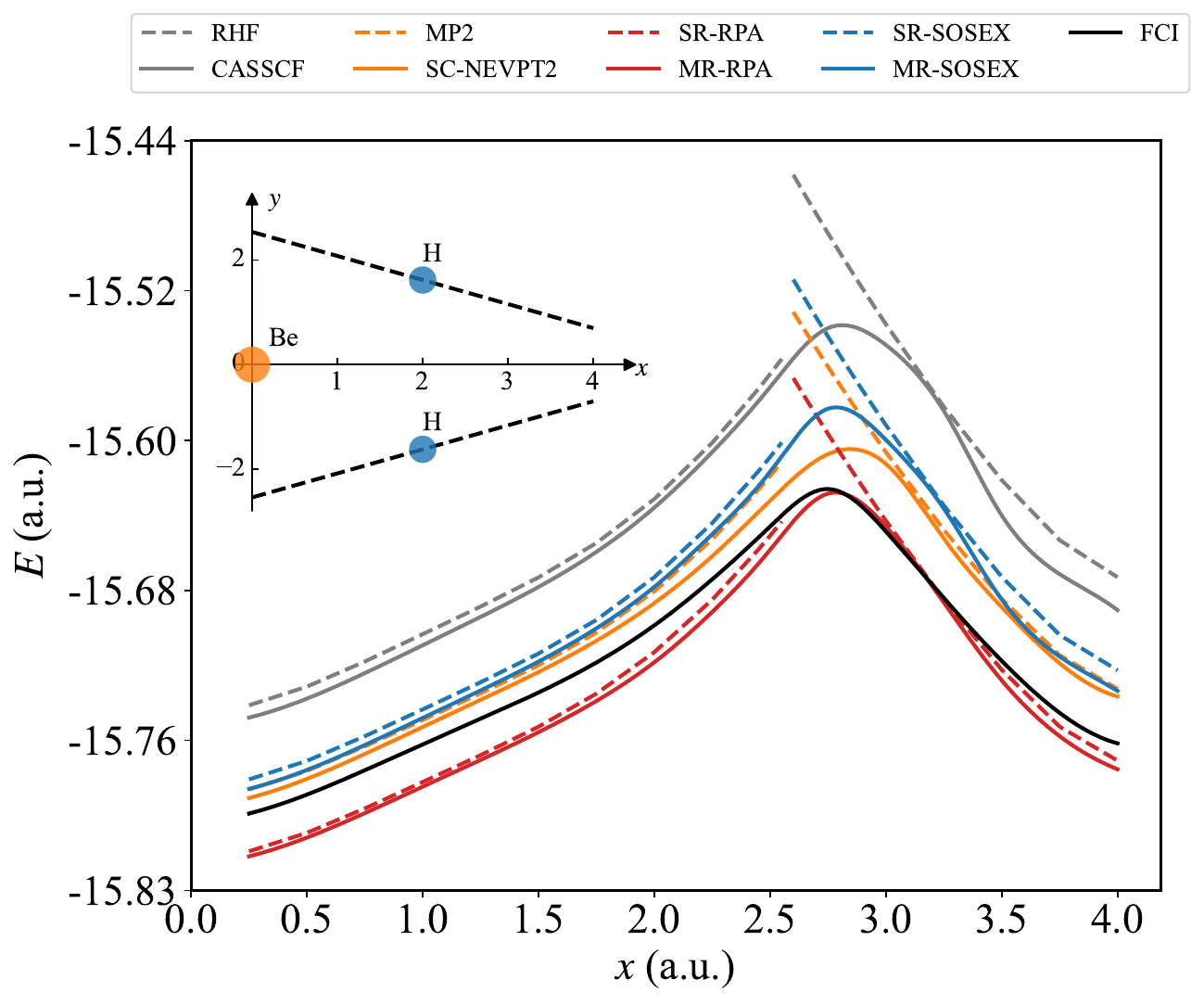}
    }
    \caption{
    Potential energy curves of the \ce{BeH2} insertion model calculated by different methods with a \ce{Be}(3s2p)/\ce{H}(2s) basis set\cite{evangelista_orbital-invariant_2011}. 
    The molecular geometry is illustrated in the inset, where a \ce{Be} atom is positioned at $(0,0,0)$ and two \ce{H} atoms are placed at $(x,\pm y,0)$ with $y = 2.54 – 0.46x$ in atomic units (black dashed lines). 
    For multi-reference methods, a CAS(2,2) active space  
    constructed with two orbitals with $b_2$ and $a_1$ symmetries, respectively, was employed. All the PECs obtained by single reference methods exhibit a jump 
    around $x=2.6$.
    }
    \label{fig:BeH2}
\end{figure}

Figure \ref{fig:BeH2} presents the computed PECs for the \ce{BeH2} insertion model, a well-established benchmark system in multi-reference studies that has been extensively investigated in previous theoretical works\cite{purvis_iii_c2_1983,kallay_general_2002,hanrath_exponential_2005,evangelista_orbital-invariant_2011,ammar2024can}. This system exhibits significant multi-reference characters in the reaction region, where the two configurations $|1a_1^2 2a_1^2 1b_2^2\rangle$ and $|1a_1^2 2a_1^2 3a_1^2\rangle$ interact strongly.
The limitations of single-reference methods are apparent in Fig. \ref{fig:BeH2}, where the use of an RHF reference results in an unphysical jump in the PEC. 
Multi-reference methods successfully address this issue. The CASSCF results exhibit an NPE of 48.1 mH, while MR-RPA, MR-SOSEX and SC-NEVPT2 demonstrate improved performance, reducing the NPE to 25.3, 35.3 and 27.4 mH, respectively.

Finally, we illustrate the size extensivity using a \ce{Li2} dimer model, consisting
of two parallel \ce{Li2} molecules (with \ce{Li-Li} bond lengths fixed at 3 \AA) 
separated by a distance of 100 \AA.
The active space is constructed from the bonding and anti-bonding orbitals of the \ce{Li-Li} bonds, forming a CAS(4,4) active space that naturally separates into two independent CAS(2,2) subspaces corresponding to each \ce{Li2} unit. Table \ref{tab:Li4} shows the computed energy for
this model and twice the energy for the individual \ce{Li2} molecule using various multi-reference methods with the cc-pVDZ basis set. The results demonstrate that both 
MR-RPA and MR-SOSEX exhibit size extensivity. This behavior originates from the fundamental property that MR-RPA/SOSEX energies exclusively contain connected diagrams, 
thereby guaranteeing proper scaling behavior with system size.

\begin{table}
    \renewcommand{\arraystretch}{1.2}
    \setlength{\tabcolsep}{2mm}
    \caption{
Size extensivity test for multi-reference methods using the \ce{Li2} dimer model. 
Energies (in Hartree) are calculated with the cc-pVDZ basis set.
    }
    \centering

    \scalebox{0.76}{\begin{tabular}{cccc}
    \hline\hline
      Method       &  2$\times E$(\ce{Li2}) & $E$(\ce{Li2}$\cdots$\ce{Li2}) & Difference \\
    \hline
      CASSCF       &  \num{-29.76044742}     & \num{-29.76044741 }   &  \num{6e-10}   \\
      MR-RPA       &  \num{-0.04806161 }     & \num{-0.04806160  }   &  \num{5e-9}   \\
      MR-SOSEX     &  \num{-0.02423804 }     & \num{-0.02423804  }   &  \num{2e-9}   \\
      SC-NEVPT2    &  \num{-0.01551936 }     & \num{-0.01551937  }   &  \num{-9e-9}  \\
    \hline\hline
    \end{tabular}}
    
    
    \label{tab:Li4}
\end{table}

{\it Conclusion.}
We have introduced a diagrammatic multi-reference generalization of MBPT.
The essential step in overcoming the lack of Wick's theorem for an interacting $\hH_0$ 
is to use the cumulant decomposition of Green's functions\cite{negele_quantum_1998}. 
This allows us to develop a diagrammatic expansion for the correlation energy
in terms of generalized Feynman diagrams composed of 
the residual interactions, zeroth-order one-body Green's functions,
and connected many-body Green's functions.
Similar to standard MBPT, the linked cluster theorem still holds, as shown in Eqs.
\eqref{eq:en} and \eqref{eq:enlinked}.
The present theoretical framework 
provides a systematic way to go beyond second-order perturbation theory via diagrammatic resummation techniques.
As a concrete demonstration of this formalism, we have derived diagrammatic multi-reference generalizations of RPA and SOSEX. Our generalization is more natural
compared to existing multi-reference generalizations of RPA\cite{yeager_multiconfigurational_1979,jorgensen_linear_1988,helmich-paris_casscf_2019,chatterjee_excitation_2012,pernal_intergeminal_2014, chatterjee_minimalistic_2016,pernal_electron_2018,pernal_exact_2018,pastorczak_correlation_2018,drwal_efficient_2022,matousek_toward_2023,guo_spinless_2024,szabados_ring_2017,margocsy_ring_2020},
as it provides a unified set of equations that seamlessly bridges the SR and MR cases.
Benchmark calculations on prototypical molecular systems demonstrate
that MR-RPA/SOSEX show significant improvements over SR-RPA/SOSEX in strongly correlated regimes.

This theoretical development opens several promising avenues for advancing \emph{ab initio} computational methods for strongly correlated systems. Future directions include the development of low-scaling MR-RPA algorithms for large molecules and materials\cite{eshuis2010fast,ren2012resolution,kaltak_cubic_2014},
the combination with density functional theory,
and the employment of alternative multi-determinant wavefunctions 
such as the generalized valence bond (GVB) state\cite{bobrowicz1977self,xu_block_2013}
as a cost-effective correlated zeroth-order wavefunction.
Notably, our diagrammatic framework provides a natural pathway for developing a multi-reference generalization of the 
$GW$ approximation\cite{hedin1965new} based on the present MR-RPA.
Progress along these directions will be reported in due time.


\begin{acknowledgement}
The authors acknowledge Yang Guo and Xinguo Ren for helpful comments on
the manuscript. This work was supported by the Innovation Program for Quantum Science and Technology (Grant No. 2023ZD0300200) and the Fundamental Research Funds for the Central Universities.

\end{acknowledgement}

\begin{suppinfo}

Detailed theoretical derivations for MR-RPA, MR-RPA matrix elements with a CASSCF reference, computational
details and additional numerical results.

\end{suppinfo}





\bibliography{main}
\end{document}